\documentclass[11pt,twoside]{article}

\usepackage{asp2006}
\usepackage{fancyhdr,graphicx,caption,rotating,multirow,lscape}
\DeclareGraphicsExtensions{.eps,.eps.gz,.ps,.jpg,.tiff}

\usepackage{natbib}
\begin{document}

\title{Asteroseismology: a powerful tool to complement planet transits}  



\author{D. Stello}
\affil{School of Physics, University of Sydney, 2006 NSW, Australia}
\author{H. Kjeldsen}  
\affil{Institut for Fysik og Astronomi (IFA), Aarhus Universitet, 8000 Aarhus,
Denmark}
\author{T. R. Bedding}
\affil{School of Physics, University of Sydney, 2006 NSW, Australia} 


\begin{abstract} 
The study of stellar oscillations - asteroseismology - has revolutionized
our understanding of the physical properties of the Sun
and similar potential for other stars has been
demonstrated in recent years. 
In
particular, asteroseismic studies can constrain the stellar size, temperature
and composition, which are important parameters to our understanding of
planetary structure and evolution. This makes asteroseismology a very
powerful tool to complement planetary transits.
As an example, the transit measurement alone does not give the radius of the
planet unless the radius of the host star is known, which again requires a
known distance to the system. Transit measurements will therefore often
require additional measurements to establish the radius of the planet. With
asteroseismology we can determine the radius of a star to very high precision
(2--3\%) using only the photometric transit measurements. This will be very
valuable for a mission such as Kepler, which will produce photometric time
series of very high quality.
\end{abstract}


\section{Introduction}   
Asteroseismology is a very powerful
tool. This is because we can measure the frequencies of the stellar
oscillations to a very high accuracy, and the oscillations can give us 
information about the stars that we cannot otherwise measure, 
helping us to constrain better the stellar parameters.  
In the last few years we have seen quite a breakthrough in results, which
indicates the great potential of asteroseismology as a tool in relation to
the search and study of exoplanets. For thorough reviews on
asteroseismology see \citet{BrownGilliland94,Dalsgaard04}

\section{Asteroseismology of solar-like stars}
\begin{figure}[t]
\centering
\begin{minipage}[c]{.49\textwidth}
\includegraphics[width=6.6cm]{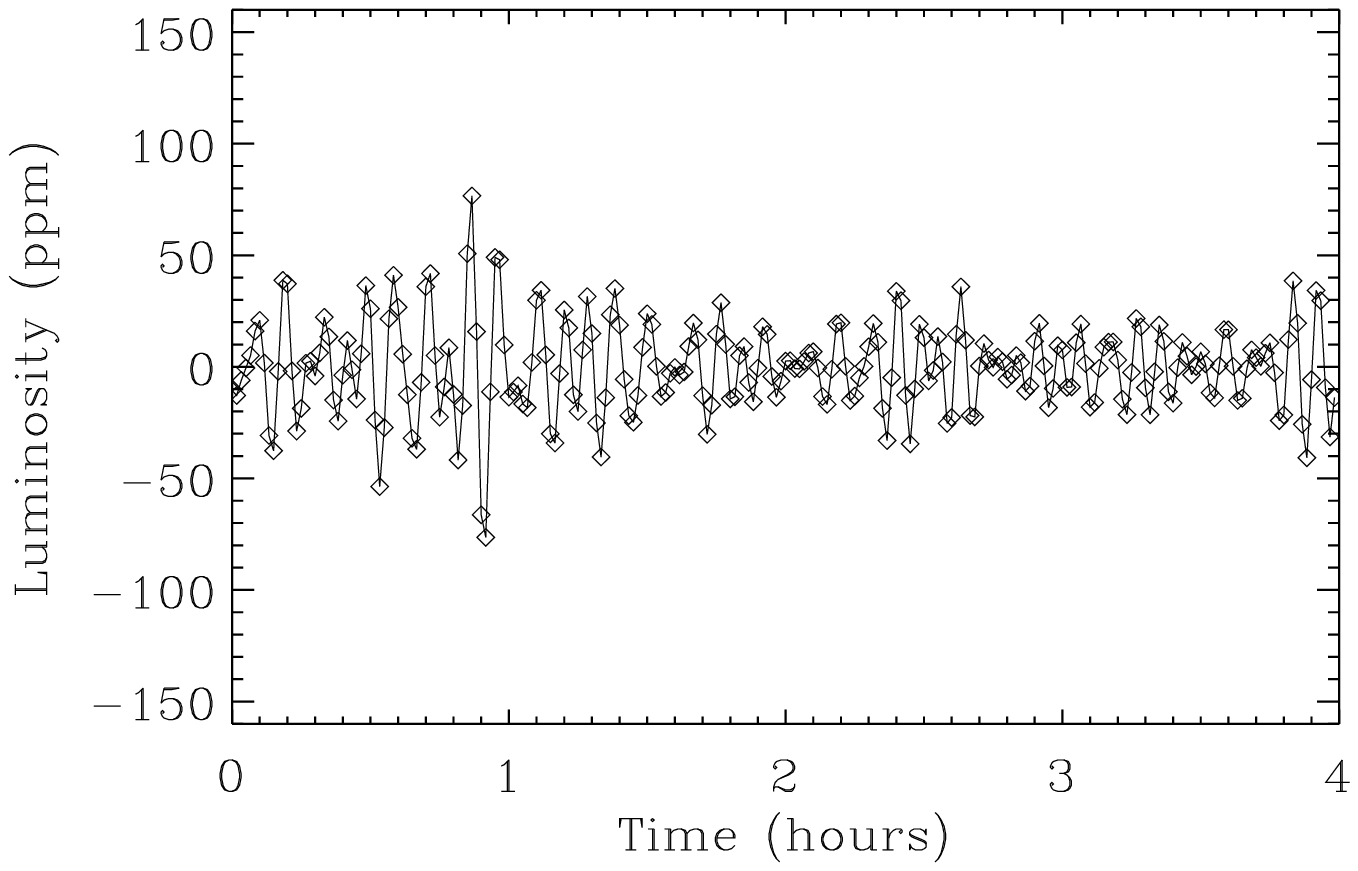}
\end{minipage}
\begin{minipage}[c]{.49\textwidth}
\includegraphics[width=6.6cm]{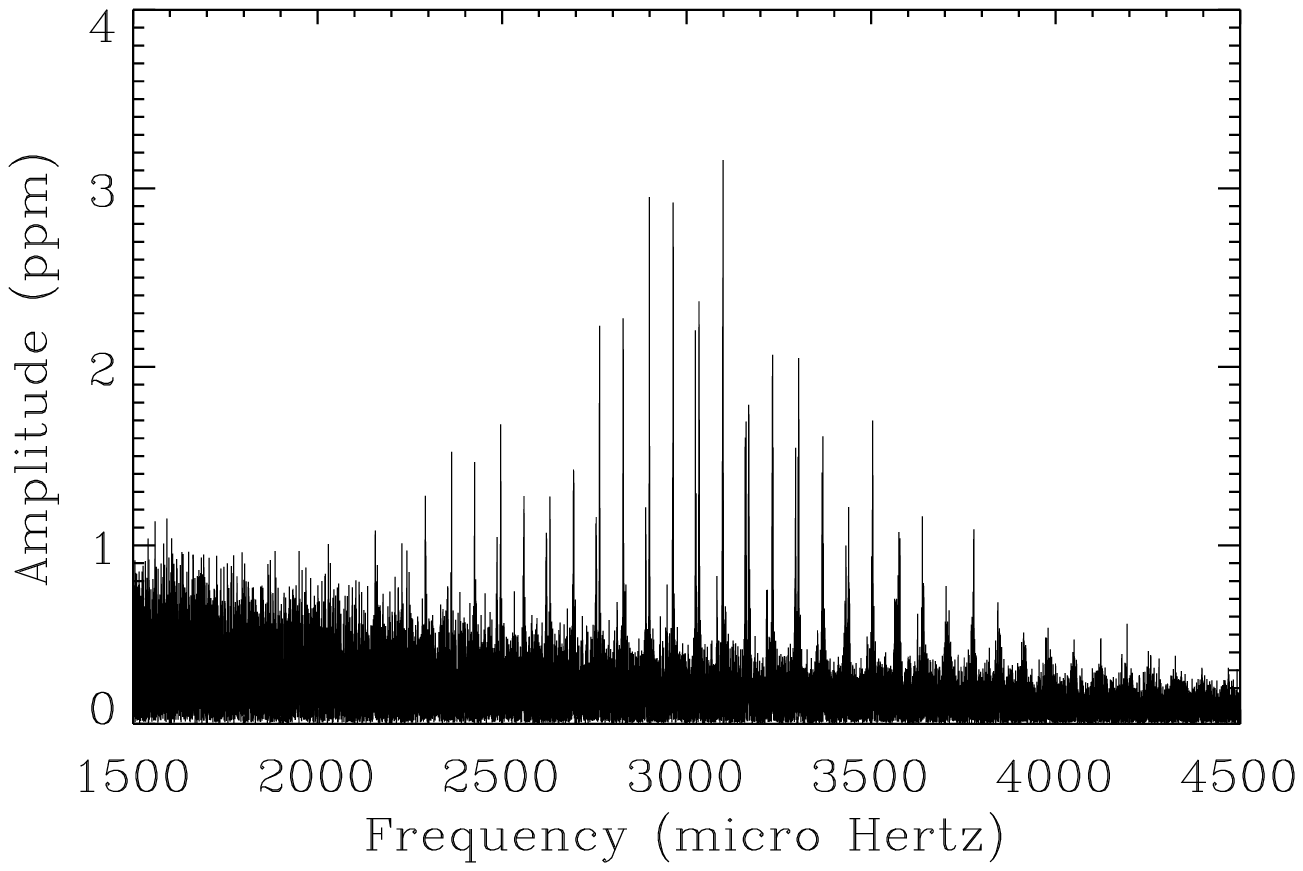}
\end{minipage}
\caption{\textbf{Left panel:} Luminosity (in parts per million) of the Sun
  observed as a star by VIRGO (green band) after granulation with
  time scales longer than $\sim9\,$minutes has been removed from the
  data. The cadence is 1 minute. 
  \label{stellofig1} 
  \textbf{Right panel:} Fourier transform of the solar time series similar to
  that seen in the left panel but including granulation at all time scales
  (60 days of VIRGO 
  observations).\label{stellofig2}} 
\end{figure}

\begin{figure}[t!]
\centering
\begin{minipage}[c]{.6\textwidth}
\includegraphics[width=8.2cm]{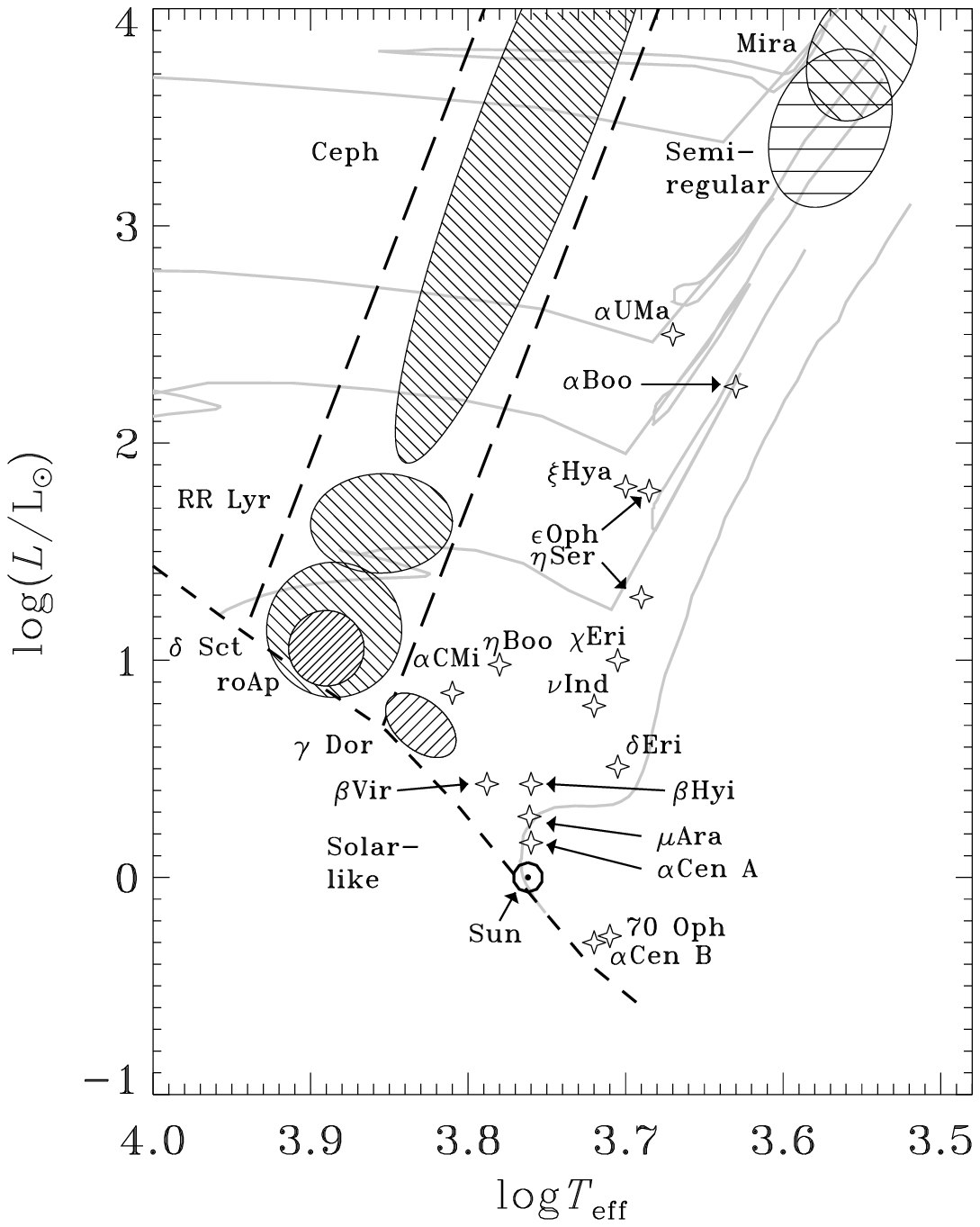}
\end{minipage}
\begin{minipage}[c]{.36\textwidth}
\includegraphics[width=5cm]{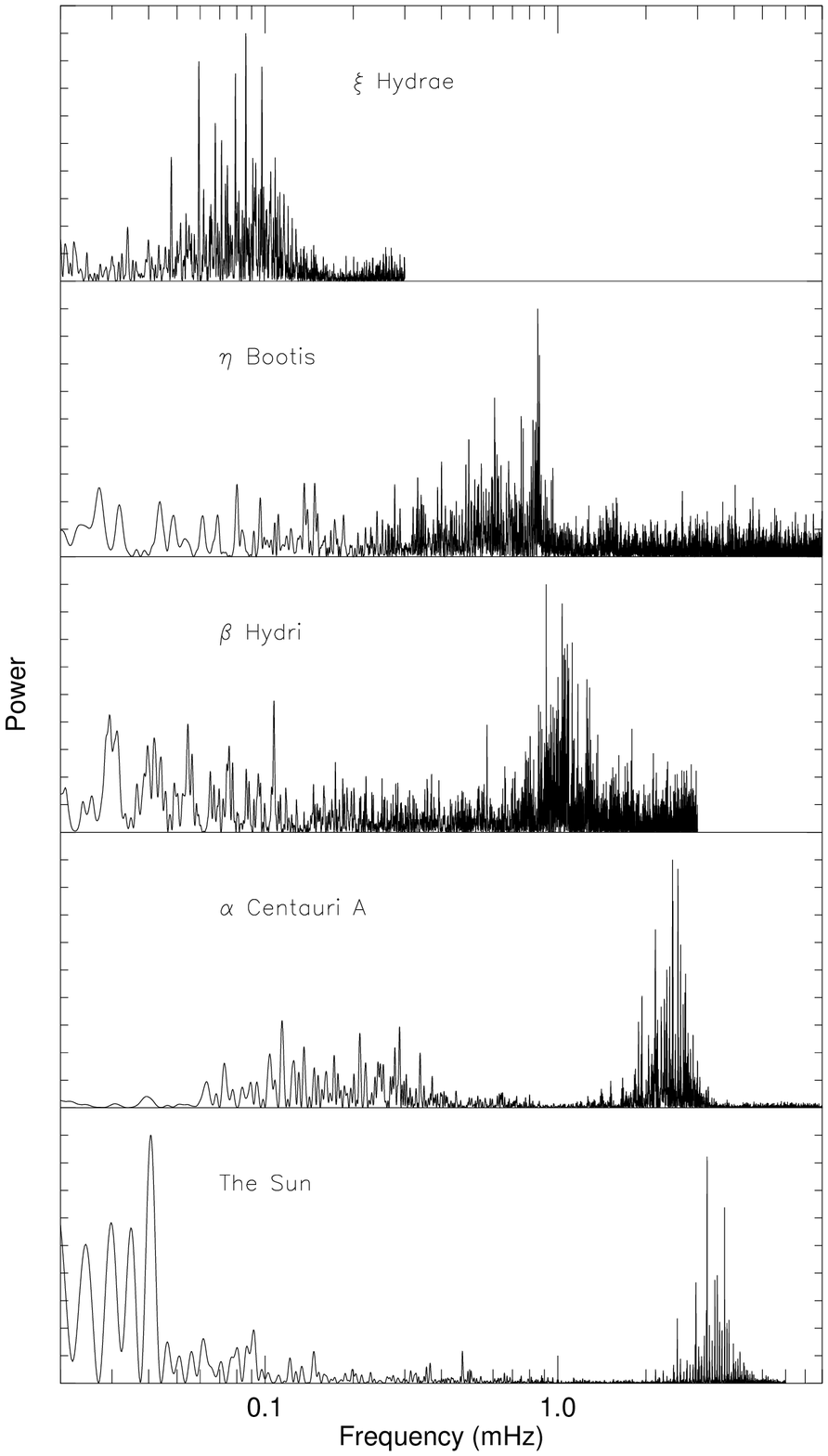}
\end{minipage}
\caption{\textbf{Left panel:} Hertzsprung-Russell diagram with names and
  approximate locations 
  of known groups of pulsating stars. The lower dashed line is
  the zero age main sequence. The two parallel long dashed lines bracket
  the instability strip. Stellar evolution tracks (grey
  curves) are shown for 1, 2, 3, 4 and 7 M$_\odot$. Stars that have shown
  evidence of solar-like oscillations are indicated. (The figure is
  slightly modified compared to the original, which was kindly supplied by
  J. Christensen-Dalsgaard.)
  \label{stellofig5}
  \textbf{Right panel:} Observed power spectra (power=amplitude squared) of
  the oscillations in the Sun and four other stars. The horizontal axis
  is the observed frequency in milli Hertz on a logarithmic scale. The
  power has been normalized to show the same peak height in 
  each plot. The actual amplitudes increase up the plot, and the peak power
  in $\xi\,$Hya is roughly 30 times that of the Sun. These results are all
  based on radial velocity measurements. For all but the Sun they are
  obtained from the ground (Sun 
  \citep{Gabriel97}; $\alpha\,$Cen A \citep{Butler04}; $\beta\,$Hyi
  \citep{Bedding01}; 
  $\eta\,$Boo \citep{Kjeldsen95}; $\xi\,$Hya
  \citep{Frandsen02}). 
  \label{stellofig6}}
\end{figure}

In Fig.~\ref{stellofig1} (left panel) 
we show the luminosity of the Sun observed as a star for four hours by the
VIRGO instrument on board the SOHO spacecraft.  
The Sun oscillates in many modes simultaneously, each mode with a slightly
different frequency, which gives the complex sinusoidal pattern
seen in Fig.~\ref{stellofig1} (left panel). 
This is clearly revealed when we take the Fourier transform of the time
series (but of longer time span). The square root this transform, also
called the amplitude spectrum is shown in Fig.~\ref{stellofig2} (right
panel).
Each peak in the amplitude spectrum represents a sinusoidal oscillation and
from this we can see that the Sun oscillates in many different modes, with
periods of roughly five minutes. The amplitudes of the oscillations are
a few parts per million in luminosity, which corresponds to temperature
changes of less than 0.01K. The oscillations can also be observed by radial
velocity measurements of the surface. The amplitude
in velocity is roughly $20\,\mathrm{cm}\,\mathrm{s}^{-1}$, corresponding
to a total displacement of the solar surface by about 100 metres.
What makes the Sun oscillate in all these modes simultaneously is 
the convection near the surface. The turbulent gas motion in the
convection zone (seen as granulation on the surface) stochastically excites 
oscillations, which are intrinsically damped, in this broad frequency
range. The rising background noise towards lower frequencies seen in
Fig.~\ref{stellofig2} (right panel) is from the random granulation cells
themselves. This noise is significantly lower in velocity (see
Fig.~\ref{stellofig6}, bottom right).

\begin{figure}[h]
\centering
\begin{minipage}[c]{.49\textwidth}
\includegraphics[width=4.5cm]{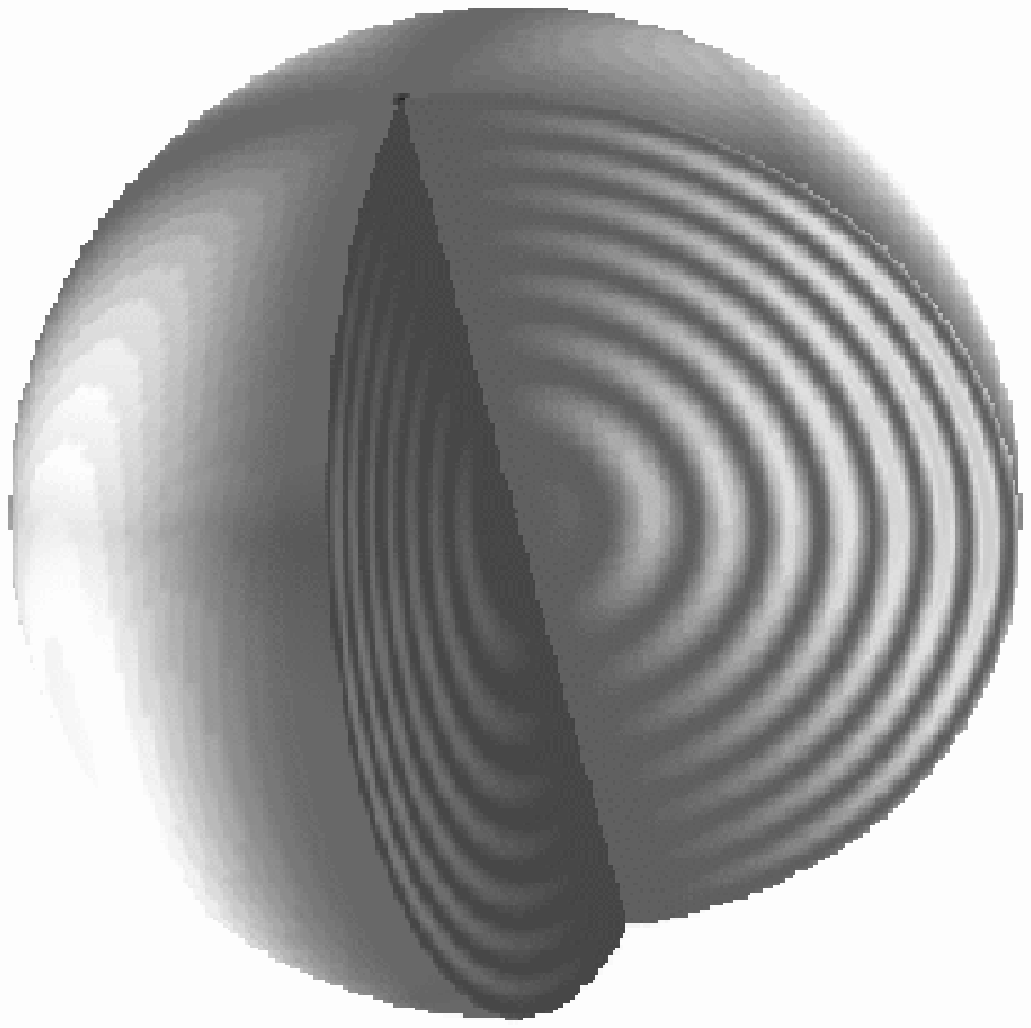}
\end{minipage}
\begin{minipage}[c]{.49\textwidth}
\includegraphics[width=6.6cm]{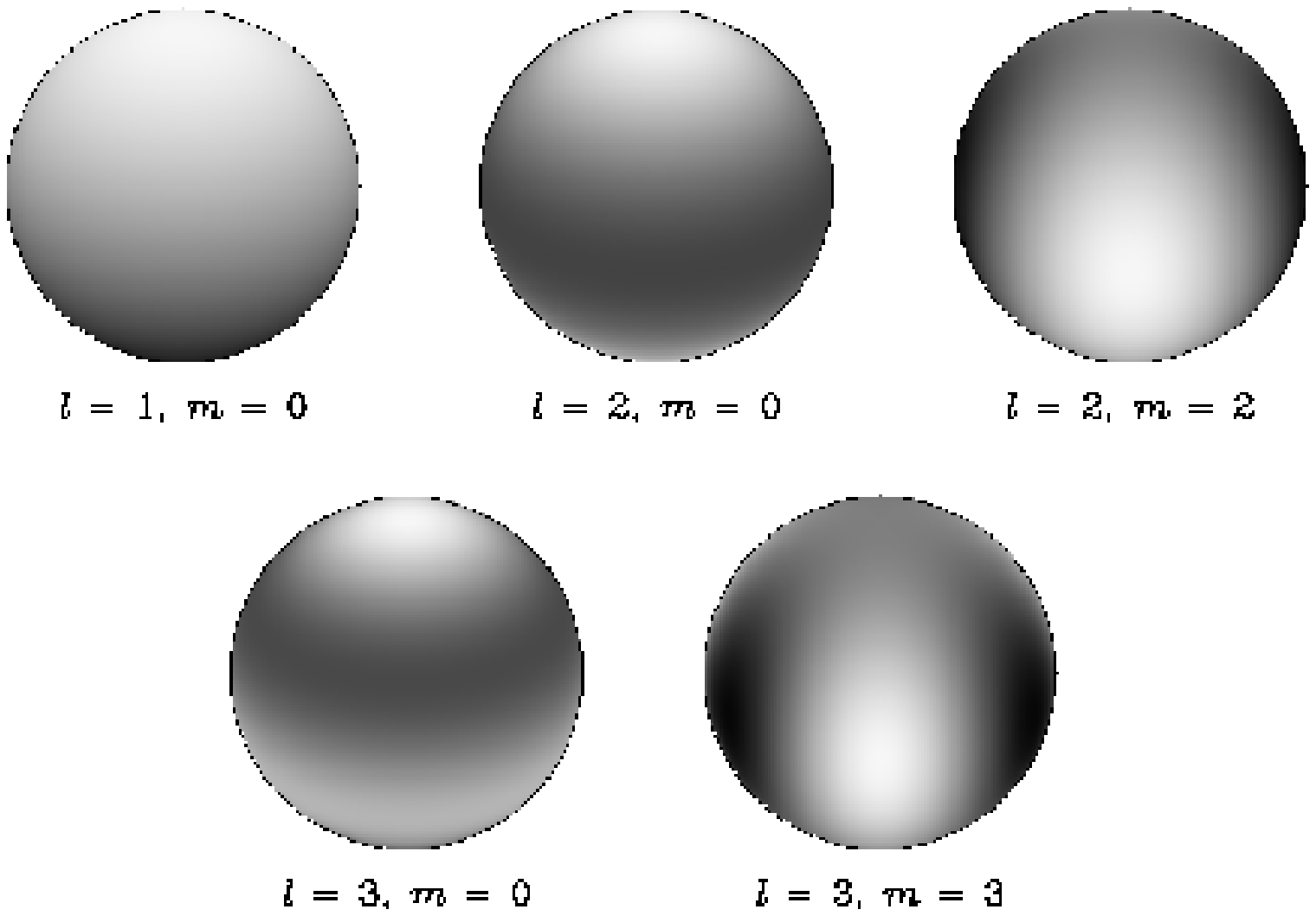}
\end{minipage}
\caption{\textbf{Left panel:} Structure of a single spherical mode with
  quantum numbers: $n=18$, $l=2$ and $m=2$ (provided by IFA, Aarhus
  Universitet). \label{stellofig3} 
  \textbf{Right panel:} Spherical harmonics representing the surface
  behaviour of the oscillations (provided by J. Christensen-Dalsgaard).
  \label{stellofig4}}
\end{figure}

Each frequency is coming from a mode of a standing sound wave, like in an
organ pipe with nodes and anti-nodes, but in three dimensions (see
Fig.~\ref{stellofig3}, left panel).  
Each mode has both a radial part described by the order $n$, which is the
number of nodes or spherical `shells', and a surface
pattern described by a spherical harmonic.
Figure~\ref{stellofig4} (right panel) shows examples of spherical harmonics
of different degree $l$ and order $m$. High-degree modes ($l>3$) cannot be
detected when the Sun is observed as a star due to geometric cancellation.  

The frequency of each mode depends on the sound speed predominantly in that
part of the Sun where the particular mode spends most of its time. By
observing the frequencies of many modes, the sound speed inside the Sun can
be reconstructed. This is extremely interesting because the sound speed
depends on physical parameters like temperature, density and
composition. For example, hydrogen turns into helium as the Sun gets
older and the sound speed and - hence the frequencies -
change. Therefore, we can 
in principle measure the age of the Sun.

A large number of stars are expected to show oscillations that are
stochastically driven by convection as in the Sun (also known as
solar-like oscillations). Stars located on the cool side of the
classical instability strip have convection near the surface, and are
therefore expected to show solar-like
oscillations. Figure~\ref{stellofig5} (left panel) shows a
Hertzsprung-Russell diagram 
where different regions of known stellar pulsations are indicated. 
In recent years solar-like oscillations have been detected in a large range
of stars from dwarfs to red giants (indicated in Fig.~\ref{stellofig5},
left panel). These results are mostly based on velocity measurements
obtained from the ground \citep{BeddingKjeldsen06}. A few examples of
these detections can be seen in Fig.~\ref{stellofig6} (right panels). The
most luminous stars (and biggest) 
oscillate at the lowest frequencies, which is clearly seen.  
From ground it is extremely difficult to measure the oscillations using
photometry due to the Earth's atmosphere, especially for the less evolved
stars, which show the smallest amplitudes. From space, however, detection
of oscillations in a true analogue of the Sun is possible.

\section{Extracting physical parameters}
The seismic signature of solar-like oscillations is a nearly regular
spaced comb pattern in the amplitude spectrum. The characteristic spacings
in this pattern can be used to characterize the star.
\begin{figure}[h]
\centering
\begin{minipage}[c]{.54\textwidth}
\includegraphics[width=7.0cm]{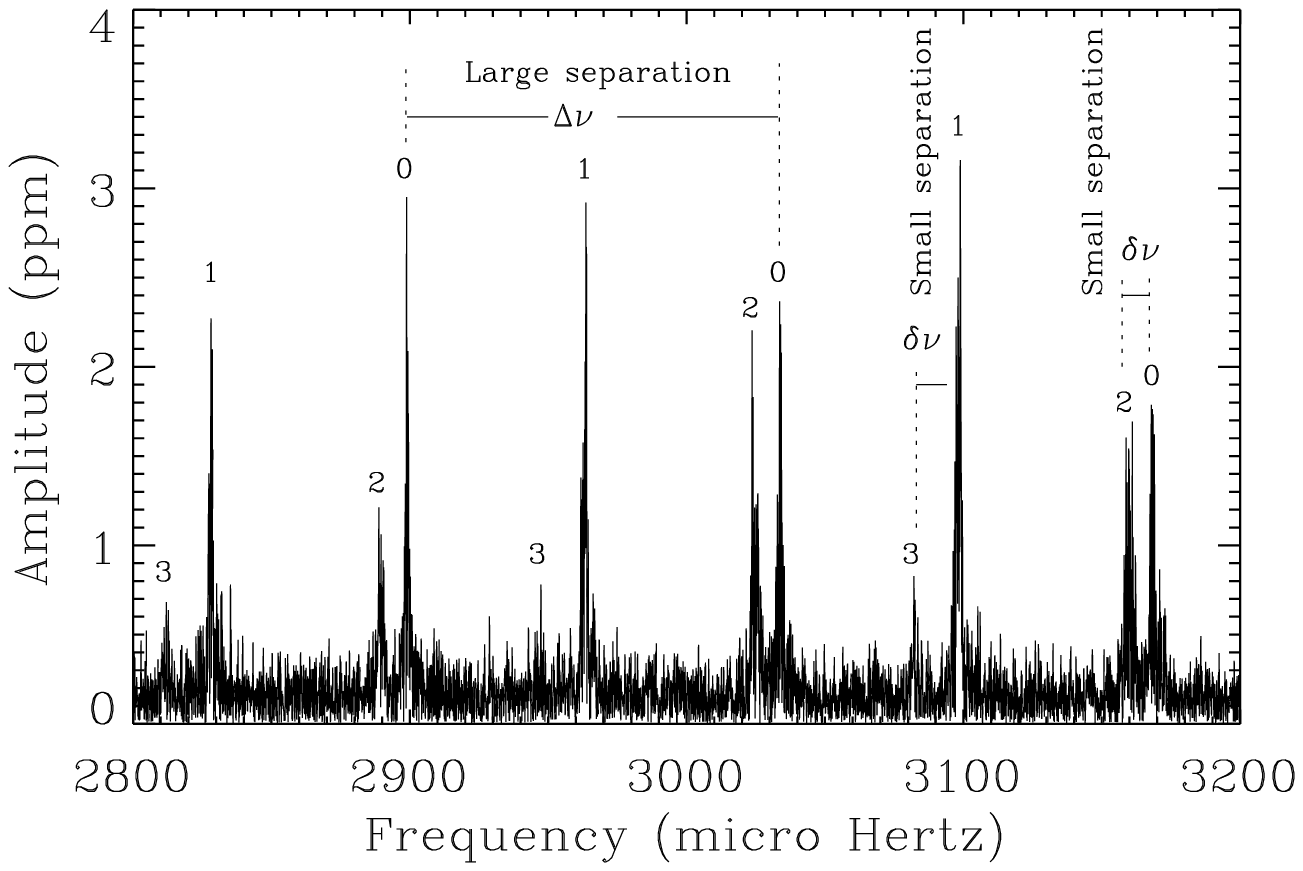}
\end{minipage}
\begin{minipage}[c]{.44\textwidth}
\includegraphics[width=6.cm]{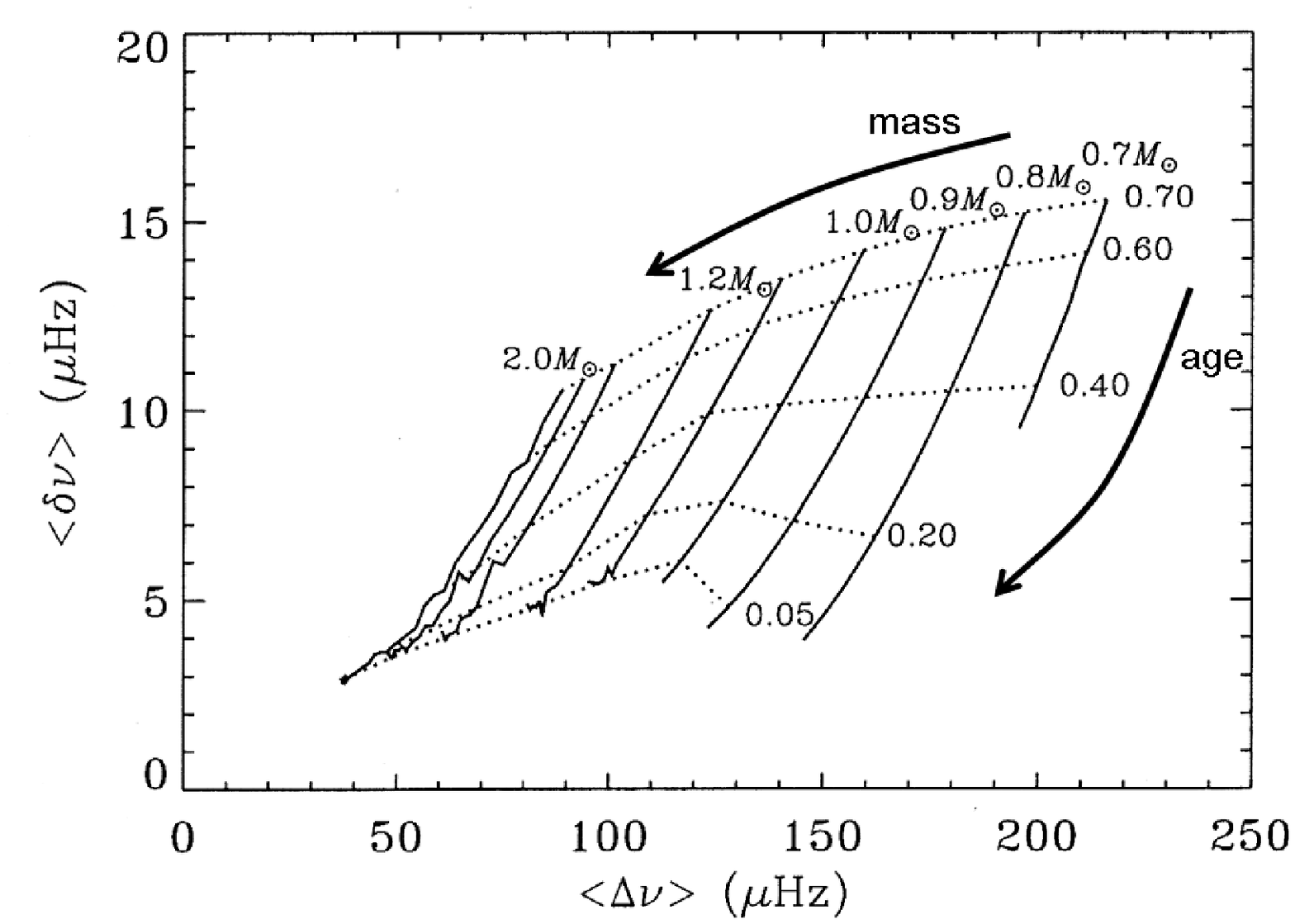}
\end{minipage}
\caption{\textbf{Left panel:} Similar to Fig.~\ref{stellofig2} but for a
  shorter frequency range for greater detail. The degree $l$ for each mode
  is indicated.\label{stellofig7}
  \textbf{Right panel:} Small versus large separation (in micro Hertz) for
  stellar evolutionary tracks (solid curves) of different mass. Dotted
  curves indicate constant central hydrogen abundance. The arrow label
  \textit{age} indicates the direction of the stellar evolution for each
  track. (The figure is a 
  slightly modified version of the original provided by
  J. Christensen-Dalsgaard.) 
  \label{stellofig8} }
\end{figure}
In Fig.~\ref{stellofig7} (left panel) we show in more detail the
amplitude spectrum 
in Fig.~\ref{stellofig2} (right panel). The degree $l$ for each mode
is indicated. We see two characteristic  
frequency separations: the large separation, which is between modes of the
same degree $l$ (but successive radial order $n$, which for the Sun is in
the order of 10--30), and the small separations.
The large separation ($\Delta\nu$) is a direct measure of the stellar
density, while the small separation ($\delta\nu$) is sensitive to the mean
molecular weight in the core (hence age) of the star. This is illustrated
in Fig.~\ref{stellofig8} (right panel), which shows stellar evolutionary
tracks of different mass (solid curves) in the $\delta\nu$ versus
$\Delta\nu$ plane. The dashed lines indicate constant central hydrogen
abundance.     

\section{Constraining the planet radius}
A planet transit potentially gives a very precise measure of the 
radius of the planet relative to its host star (Holman et al., this
proceedings). However, additional observations are required to establish
the actual size scale. Basically, we cannot tell if it is a  
transit of a big star by big planet, or a transit of a small star by a
small planet. Seismology can be of significant help to constrain the actual
stellar radius (and hence the radius of the planet when combined with the
transit depth), using only the photometric transit data. The requirements
are that the precision is high enough and that the data sample the
oscillations. These requirements depend very much on where in the
HR diagram the star is located, as both oscillation amplitude and period
vary significantly across the HR diagram. 

Without applying asteroseismology and without a distance measurement, the
stellar radius has to be constrained 
by spectroscopy and photometry. In this case we can determine the temperature
to roughly 100 K (or 2\%), and the absolute luminosity to approximately
40\%. The accuracy in the radius is therefore
roughly 20\%. From the stellar location in the HR diagram we can further
obtain a mass estimate to about 10\% using stellar models.

From seismology we get the large separation to a very high accuracy, at
least 0.5\% for Kepler targets. Since the mean density of the star is
proportional to the large separation squared, this seismic measure gives
the density to about 1\%. 
Combining that with the 2\% temperature determination reduces the
uncertainty in the mass to roughly 5\% (in addition to a smaller uncertainty
in the luminosity of approximately 10\%). This mass estimate together with the
seismic density measurement finally provides a 2--3\% radius estimate,
which is a tremendous improvement.

For Kepler, parallax measurements are expected to be
derived at the end of the mission, which  will lower the
uncertainty in the luminosity to about 2\% from the parallaxes alone. This
implies a better mass estimate to roughly 3\%, and hence a radius estimate
of approximately 1\% if combined with the seismic measurement of the large
separation. This in turn improves the precision of the temperature 
to less than 1\%. 
In addition, the small separation can be measured to at least 10\%, which
for a solar mass star implies that the age will be known to better than 1
Gyr.

\section{Conclusion}
\begin{itemize}
\item Including seismology to determine the stellar parameters improves
  significantly the accuracy with which we can characterize the stars, and
  hence their planets.  
\item This can be obtained without complicated and time-consuming
  calculations of pulsation models for each star.  
\item Only the data already obtained to detect the transits need to be
  used, provided the data sampling is high enough. Cadence requirements are
  about once per minute for a true Sun-like star, and up to roughly half an
  hour for a giant star.   
\item This technique would help tremendously for a mission such as Kepler,
  where the precision in the radius estimates of the host stars are
  improved  by a factor of 7  without any parallax information and by a
  factor  of 20 when parallaxes become available.  
\end{itemize}
In the last few years, the field of asteroseismology has taken advantage
of the observational techniques developed in the quest of detecting and
studying exoplanets. This has mostly been from ultra-high precision Doppler
measurements, but in the coming years very accurate photometry from
space will 
become available. With these observational techniques in common, very
similar data sets and complementary but supporting science goals, there is
an  obvious opportunity for exoplanet research to take advantage of the 
extremely precise measurements that seismology can offer in order to
determine the properties of planet-hosting stars. This makes
the future for both research fields very exciting.

\acknowledgements 
This paper has been supported by the Astronomical Society of Australia and
by the Australian Research Council. We are grateful for the VIRGO data
being publicly available.
%
%

\end{document}